\documentclass[final,5p,times,twocolumn]{elsarticle}

\usepackage{graphics,graphicx,dcolumn,bm,fleqn,epic,eepic,float,ulem}
\usepackage{amssymb,amsmath}
\usepackage{color}

\definecolor{red}{rgb}{1,0,0}
\definecolor{green}{rgb}{0,1,0}
\definecolor{blue}{rgb}{0,0,1}

\newcommand{\f}{\mathbf}
\newcommand{\txa}{\tilde{x}_1}
\newcommand{\txb}{\tilde{x}_2}
\newcommand{\ttxa}{\hat{x}_1}
\newcommand{\ttxb}{\hat{x}_2}
\newcommand{\no}{NO$_2$}

\journal{arxiv}
\begin{document}

\begin{frontmatter}

\title{Searching for optimal variables in real multivariate stochastic 
       data}

\author{F.~Raischel$^{a}$}
\author{A.~Russo$^{b}$}
\author{M.~Haase$^{c}$}
\author{D.~Kleinhans$^{d,e}$}
\author{P.G.~Lind$^{a,f}$}

\address{$^a$Center for Theoretical and Computational Physics, 
         University of Lisbon, Av.~Prof.~Gama Pinto 2, 
         1649-003 Lisbon, Portugal}
\address{$^b$Center for Geophysics , IDL, 
         University of Lisbon
         1749-016 Lisboa, Portugal}
\address{$^c$Institute for High Performance Computing, University of Stuttgart,
         Nobelstr.~19, D-70569 Stuttgart, Germany}
\address{$^d$Institute for Biological and Environmental Sciences,
         University of Gothenburg,
         Box 461, SE-405 30 G\"oteborg, Sweden}
\address{$^e$Institute of Theoretical Physics, University of
         M\"unster, D-48149 M\"unster, Germany}
\address{$^f$Departamento de F\'{\i}sica, Faculdade de Ci\^encias 
         da Universidade de Lisboa, 1649-003 Lisboa, Portugal} 

\begin{abstract}
By implementing a recent technique for the  determination of stochastic 
eigendirections of two coupled stochastic variables, we investigate 
the evolution of fluctuations of \no\ concentrations at two monitoring
stations in the city of Lisbon, Portugal.
We analyze the stochastic part of the measurements recorded at the monitoring stations by means of a 
method where the two concentrations are considered as stochastic variables 
evolving according to a system of coupled stochastic differential equations. 
Analysis of their structure allows for transforming the set of measured 
variables to a set of derived variables, one of them with reduced 
stochasticity.
For the specific case of \no\ concentration measures, the set of derived
variables are well approximated by a global rotation of the original set 
of measured variables.
We conclude that the stochastic sources
at each station are independent from each other and typically have
amplitudes of the order of the deterministic contributions.
Such findings show significant limitations when predicting 
such quantities.
Still, we briefly discuss how predictive power can be increased in general
in the light of our methods.
\end{abstract}


\end{frontmatter}

\maketitle

\tableofcontents



\section{Introduction}
\label{sec:int}

The industrial and urban development during the last decades has 
led to a general decrease of air quality, drastically affecting urban 
environmental and human life quality. 
Although, according to the European Environment Agency report\cite{europeanreport}, air quality has improved in general during the last years, this enhancement
was not significant enough to ensure good air quality in all urban areas.
One of the pollutants  with negative 
impact on health and environment is \no.  
Anthropogenic \no\ is mainly emitted by vehicles and industrial processes.
\no\ has not only severe effects on health causing e.g.\ respiratory and 
cardiovascular diseases, it also affects the 
environment\cite{diseases1} as nitrogen deposition leads to 
eutrophication\cite{diseases2}.
A better understanding of the mechanisms that influence  
production,  transport, and  decomposition of \no\ is therefore important. 
Previous studies revealed that temperature, wind speed and direction, relative humidity, 
cloud cover, dew point temperature, sea level pressure, precipitation, and 
mixing layer height are relevant meteorological variables to model the concentrations of air 
pollutants\cite{diseases1,mechanismsNO2-1,mechanismsNO2-2,mechanismsNO2-3,mechanismsNO2-4}.  
In particular, approaches that deal with the evolution of the
\no\ concentration at individual city locations are important for forecasting
the air quality of urban regions.
\begin{figure}[t]
  \centering
  \includegraphics[width=0.5\textwidth]{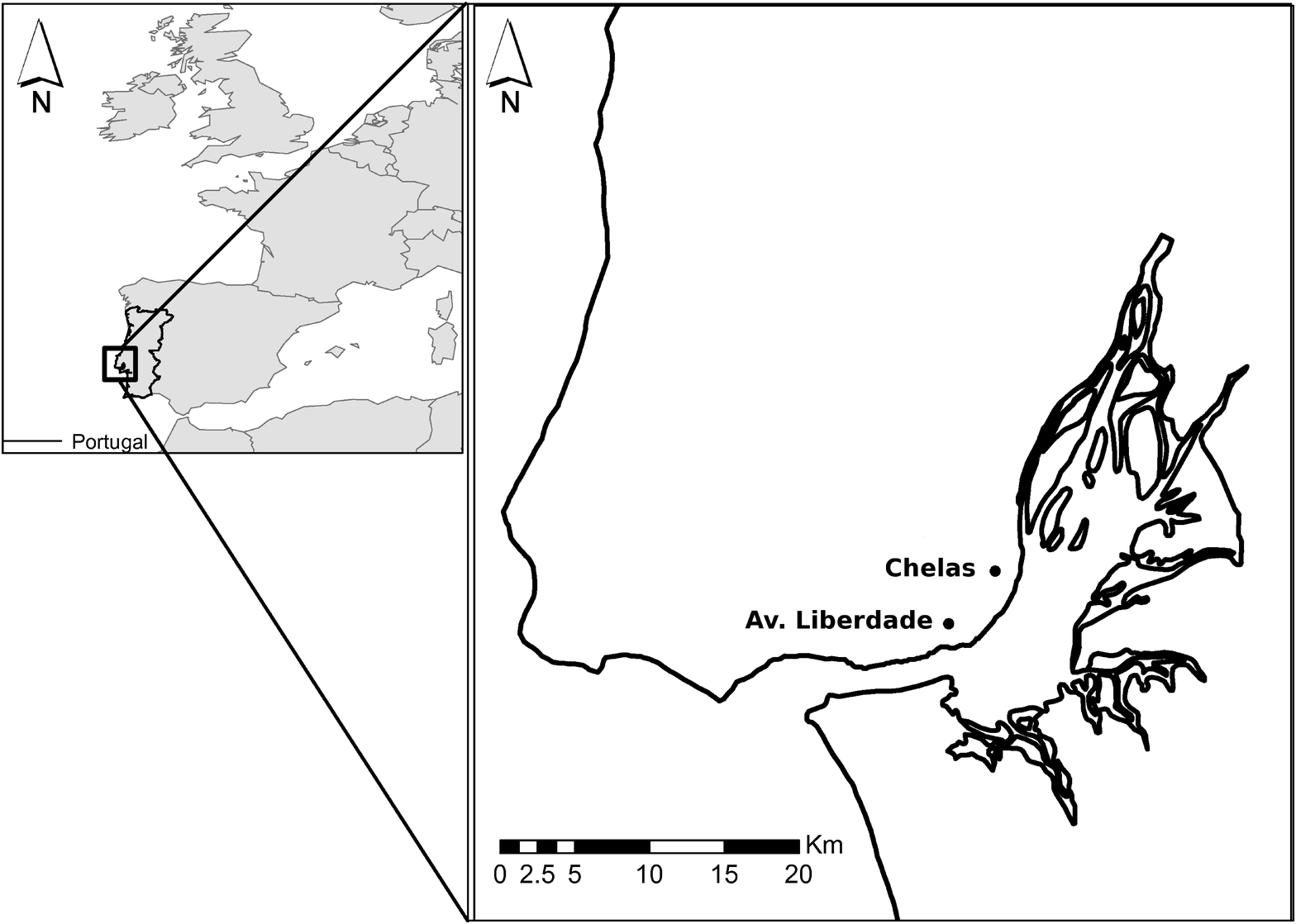}
  \caption{\protect
           \no\ measurement stations in the region of Lisbon (Portugal)
           at the Southwestern coast of Europe.
           In this paper we focus on the set of measurements taken at the
           stations of Chelas and Avenida da Liberdade with
           approximately $10^5$ data points each extracted in the
           period between 1995 and 2006.} 
\label{fig1}
\end{figure}

Recently a  framework\cite{friedrich97,physrepreview} for 
analyzing measurements on complex systems was introduced, aiming for 
a quantitative estimation of drift and diffusion functions from measured 
data. These functions can be identified with the deterministic and stochastic 
contributions to the dynamics, respectively, and give a considerable insight 
into the underlying systems. 
The framework was already successfully applied for instance
to describe turbulent flows\cite{friedrich97} and the evolution
of climate indices\cite{lind05,lind07}, stock market indices\cite{friedrich00}, 
and oil prices\cite{ghasemi07}. At the same time, the basic method has been refined 
in particular with respect to the impact of finite sampling effects 
\cite{kleinhans05,lade09}, the impact of measurement 
noise\cite{boettcher06,lind10,carvalho2010}, and the role of local 
eigendirections of the diffusion matrices \cite{vitor}.

In this paper, we aim to apply some recent methods for deriving variables
with reduced stochastic fluctuations to empirical data.
Namely, we  adapt this framework  for analyzing  measurements of 
\no\ concentrations in the 
metropolitan region of Lisbon, Portugal
(see Fig.~\ref{fig1}), taken over several years. 
We argue that the temporal fluctuations of these concentrations 
result from two independent 
contributions: one periodic and one stochastic.
The periodic part describes daily, weekly, seasonal and 
yearly variations of the concentration, which is an accepted and 
well-studied result\cite{periodicpartNO2}.
The stochastic contribution can be modelled through a stochastic differential
equation\cite{fpeq} having two terms, one drift forcing (deterministic) and one
diffusive fluctuation (stochastic).

When addressing stochastic higher-dimensional systems it is typically 
difficult to identify the variables most relevant for a proper description 
of the system's evolution.
In geophysical applications, the reduction of the full set of variables
 to only  a few variables often is achieved by means of the
so-called Principal Component Analysis (PCA)\cite{pca}
or other standard reduction methods, such as stepwise regression or 
ARIMA\cite{Wilks2006}.
However, the inherent fluctuations are not so commonly investigated.

In this paper we will apply a recent method for reconstructing the phase
space of two stochastic variables, which evolve according to a set of
two coupled stochastic equations defined through drift vectors and
diffusion matrices \cite{vitor}. The method is based on the eigenvalues 
of the diffusion matrices, from which it is possible to derive a path in 
phase space through which the deterministic contribution is enhanced. 
This technique implies a transformation of variables and 
allows for the investigation of  the minimal number 
of independent sources of stochastic forcing in the system.
In particular, a rather small eigenvalue of the diffusion matrix, compared 
to the average value of all the other, corresponds to one eigendirection 
in which stochastic fluctuations may be neglected, reducing the number
of stochastic variables taken for describing the system's evolution.
On the contrary, having all eigenvalues of the same order of magnitude
means that the number of independent stochastic forces equals the
number of variables.
Moreover, as we will see, a direct inspection of the diffusion
functions enables one to ascertain if the stochastic contributions,
one for each variable, are coupled among them or not.
Therefore, we argue that the diagonalization of the diffusion matrices 
gives insight into the system.
\begin{figure}[tb]
  \centering
  \includegraphics[width=0.5\textwidth]{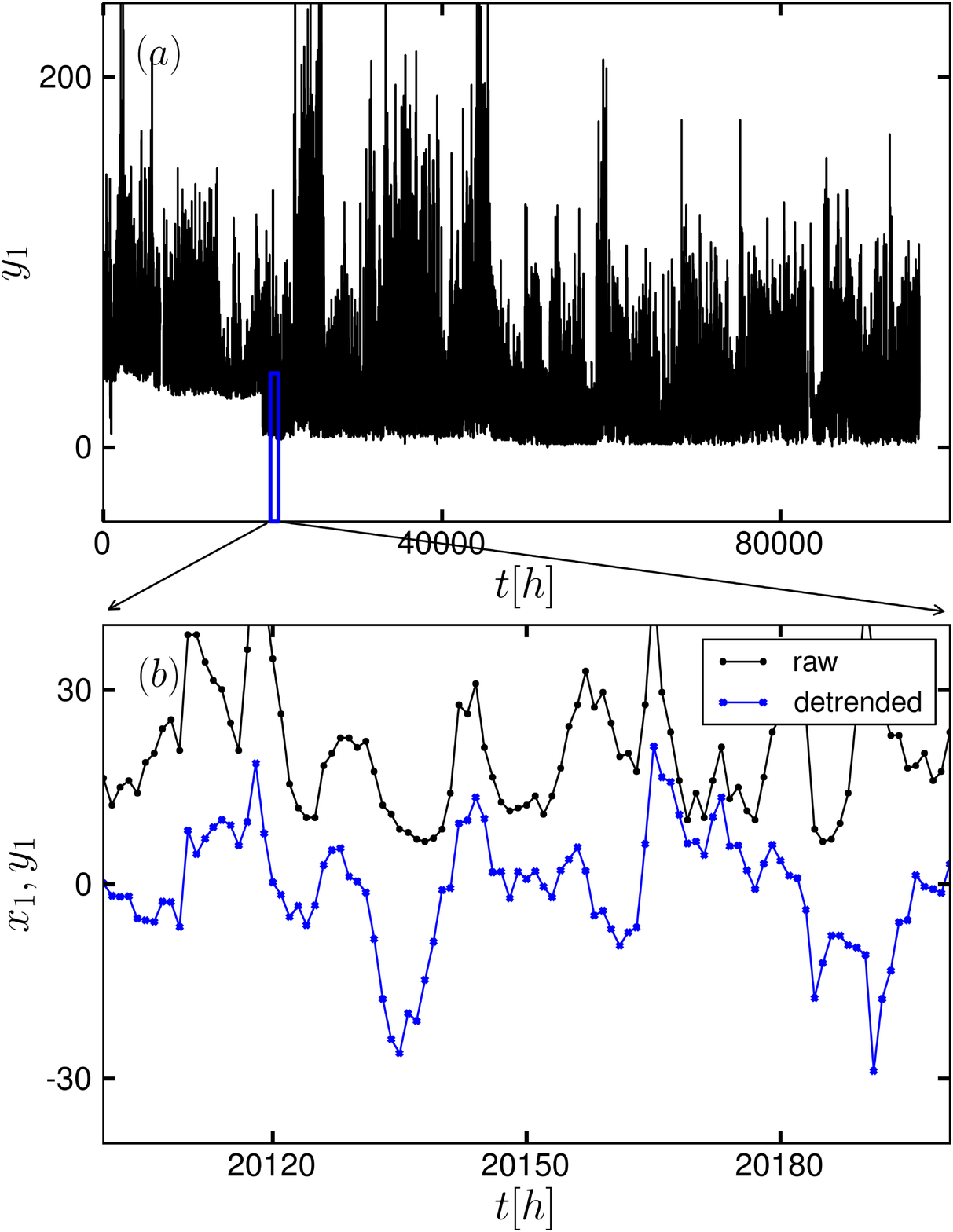}
  \caption{\protect 
     {\bf (a)} Time series of the \no\ concentration at the
     station of Chelas, before detrending according to
     Eq.~(\ref{eq:detrend}), and
     {\bf (b)} a zoom-in of these ``raw'' time series $y_1$  compared to the
     detrended series $x_1$, 
     which takes averages of $52$-weeks periods, and then a second
     detrend with daily averages.
     Vertical offset of same plots are done for clarity.
     For the station at Avenida da Liberdade similar features are found
     (not shown).}
\label{fig2}
\end{figure}

We start in Sec.~\ref{sec:data} by describing the properties and the 
preparation of the data set.
Consecutively, in Sec.~\ref{sec:method} and \ref{sec:markov}, 
the modeling of the time series as a Langevin process is carried out and 
its transformation to a new coordinate system are described in 
Secs.~\ref{sec:eigensyst} and \ref{sec:transform} respectively.
In Sec.~\ref{sec:compari} we discuss the performance of the transformation 
of the coordinates obtained by our approach compared to other techniques 
commonly used for statistical analysis of measured data. 
Section \ref{sec:results} closes this Letter with a general summary and 
ideas on the interpretation of the transformed time series with respect to 
the underlying  environmental processes.

\section{\no\ measurements in Lisbon}
\label{sec:data}

In this section we briefly describe the sets of data analyzed in this paper
as well as its preparation for analyzing the stochastic components of the 
measurements.

The data set covers hourly measurements of \no\ concentration, 
taken at 22 stations in the urban center of Lisbon
recorded from of 1995 to 2006. 
For this study we choose the data from 1995 to 2005 for the monitoring
stations at Chelas and at Avenida da Liberdade.
These stations are located  at a distance of 
$\sim 4\ km$ from each other, see Fig.~\ref{fig1}. 
In the following, the \no\ concentrations at the stations of Chelas and
Avenida da Liberdade will be designated  as $y_1(t) $ and $y_2(t)$, 
respectively, omitting the temporal dependency when not necessary. 

Increments in time are always of $1$ hour.
Each of the data sets contains $10^5$ measurement points approximately,
including some periods of incomplete or erroneous measurements that are 
disregarded for our analysis.
In the case of the chosen stations, the series of measurements $y_1$ and  
$y_2$ contain 3726 and 4548 instances of measurement errors, respectively.

The concentration of \no\ is strongly driven by daily,
 weekly, monthly and yearly anthropogenic routines, and also by
periodic atmospheric processes. 
For instance the rush hours on working days have an almost immediate 
impact on the \no\ concentration, and thus, on air quality.
The 24 hours and one week cycles  are both traffic related and mirror daily 
and weekly cycles.
The measurements of \no\ are therefore influenced by different periodic
forcings and, since we are interested in the fluctuations of \no\ 
concentrations, the periodic behavior must be first detrended.
The detrended series for $y_1$ and $y_2$, represented below as
$x_1$ and $x_2$ respectively, are obtained as follows.

One first partitions
the data in segments of length $N$, which we suppose to be a multiple
of relevant periodic fluctuations in the data set. As a second step, a
mean segment is calculated by averaging measurements with the same
position in the segment over the entire data set according to
\begin{equation}
  \mathcal{N}_i(n)\equiv\left\langle y_i(t)|t=n+mN, m=0,1,\ldots\right\rangle\quad
  \label{average}
\end{equation}
for $n=0,1,...,N-1$. The detrended data set $x_i$ is then calculated by 
subtracting the respective values of the mean segment from the measured 
data, 
\begin{equation}
  x_i(t)\equiv y_i(t)-\mathcal{N}_i(t\,\mathrm{mod}\, N)\quad .
  \label{eq:detrend}
\end{equation}
for $t=1,\dots,T$ with $T>N$.
If $T$ is the size of the data set, our simulations have shown that 
averages over $N=52$ weeks is the best choice  for the entire
data set, to take 
into account all known periodicities mentioned above.
With this detrending method, some periodicities with variable phase remain.
To filter also these periodicities, a second detrending with 
$N=1$ day is then performed on consecutive periods of $T=14$ days.

Figure \ref{fig2}a shows the original data $y_1$ for the station of Chelas.
A zoom-in of a small time interval is plot in Fig.~\ref{fig2}b together
with the corresponding detrended data $x_1$.
From now on, 
if not stated explicitly otherwise, we will only consider the detrended 
time series $x_1$ and $x_2$. 
Next describe their characteristics by means of a stochastic process.

\section{Modeling stochasticity in series of \no\ concentrations: 
         Langevin processes}
\label{sec:method}

The detrended series $x_i$ in Eq.~(\ref{eq:detrend}) reflect the remaining 
stochastic components of the measurements at the respective stations of 
Chelas and Avenida da Liberdade.
In this section we assume that, with two variables, the stochastic process 
is modeled by a system of two coupled Langevin equations,
containing a deterministic and a stochastic part, described through a 
drift vector and a diffusion matrix, respectively.
\begin{figure}[htb]
  \centering
  \includegraphics[width=0.5\textwidth]{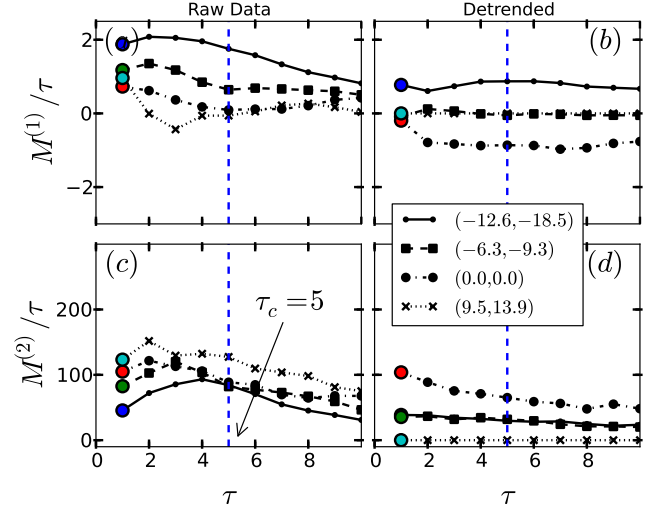}
  \caption{\protect
     First conditional moments $M^{(1)}$ for 
     {\bf (a)} the original series and
     {\bf (b)} the detrended series, with different \no\ concentrations 
     $(x_1,x_2)$
     at each one of the two stations (see legend).
     The corresponding second conditional moments $M^{(2)}$ are 
     shown in {\bf (c)} and {\bf (d)}, respectively. 
     These moments are computed according to 
     Eqs.~(\ref{M1}) and (\ref{M2}).
     While the original data presents oscillations beyond a
     given time interval $\tau_c\sim 5$, the detrended time
     series does not (see text). The value of the corresponding
     Kramers-Moyal coefficient at the value of $(x_1,x_2)$ chosen
     is given by Eq.~(\ref{DefCoefKM}) for the lowest value
     of $\tau$, i.e.~one.}
\label{fig5}
\end{figure}

For the general case of a $K$-dimensional state vector  
${\bf X}=(x_1,...,x_K)$, the It\^o-Langevin equations describing the 
evolution of a particular trajectory in time read
\cite{fpeq,gard}:
\begin{equation}
\frac{d \mathbf{X}}{d t}= 
                     \mathbf{h}(\mathbf{X})
                     + \mathbf{g}(\mathbf{X}) 
                     \mathbf{\Gamma}(t) , 
\label{Lang2DVect}
\end{equation}
where $\mathbf{\Gamma}=(\Gamma_1,\dots,\Gamma_K)$ is a set of $K$ independent
stochastic forces with Gaussian distribution fulfilling 
\begin{subequations}
\begin{eqnarray}
\langle \Gamma_i(t)\rangle &=& 0 \label{meangamma}\\
\langle \Gamma_i(t)\Gamma_j(t')\rangle &=& 2\delta_{ij}\delta(t-t') .
\end{eqnarray}
\label{stdgamma}
\end{subequations}
The two terms on the right hand side of Eq.~(\ref{Lang2DVect})
include both the deterministic contribution, $\mathbf{h}=\{ h_i\}$, 
and the stochastic contribution, $\mathbf{g}=\{ g_{ij} \}$.
The deterministic contribution 
describes the physical forces which drive the system,
while functions $\mathbf{g}$ account for 
the amplitudes of the different sources of fluctuations 
$\mathbf{\Gamma}$\cite{physrepreview}.
\begin{figure*}
  \centering
  \includegraphics[width=0.99\textwidth]{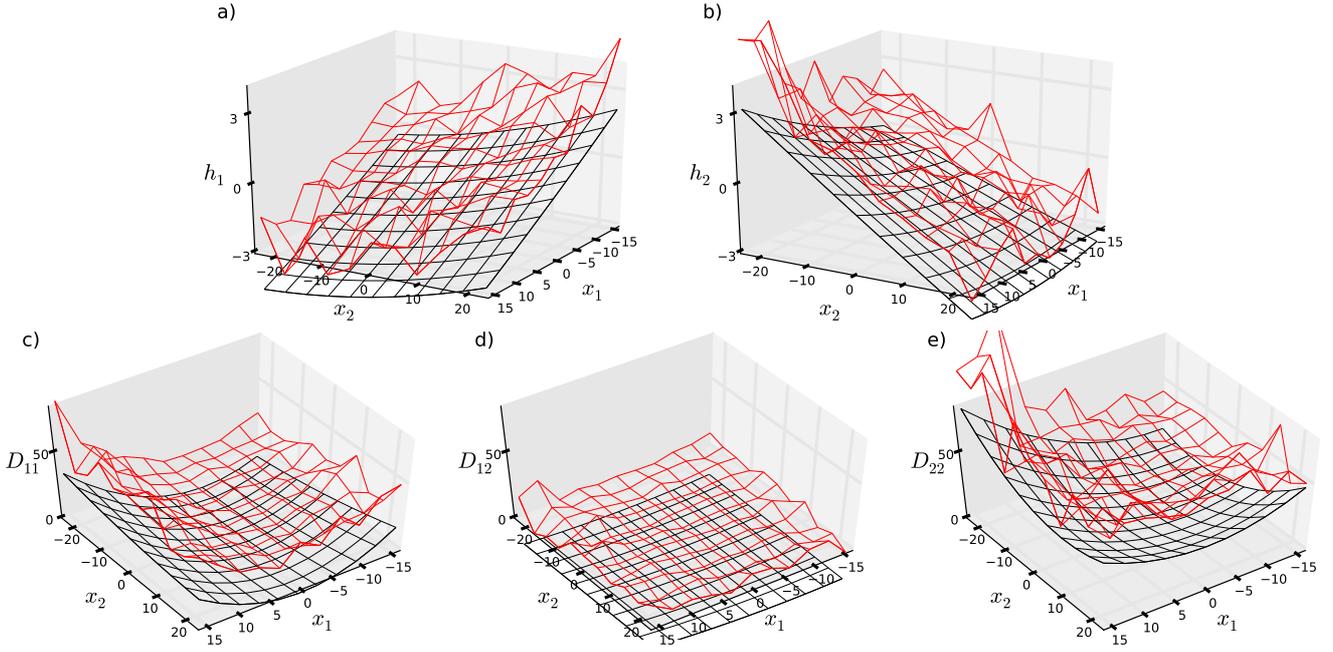}
  \caption{\protect 
      For the detrended series we plot
      {\bf (a-b)} both components of the 
      drift vector $\mathbf{h}=(h_1,h_2)$ and 
      {\bf (c-e)} the components of diffusion matrix 
      $\mathbf{D}^{(2)}=\{ D^{(2)}_{ij}\}$.
      The corresponding fitted surfaces (black) are vertically 
      offset for clarity.
      Since $\mathbf{D}^{(2)}$ is symmetric (see text) one has 
      $D^{(2)}_{12}=D^{(2)}_{21}$.}
\label{fig6}
\end{figure*}

The coefficients $\mathbf{h}$ and  $\mathbf{g}$ are directly related to
the drift vectors and diffusion matrices\cite{fpeq} 
\begin{eqnarray}
D_i^{(1)}(\mathbf{X}) &=& h_i (\mathbf{X}) \label{Deqh1}\\
D^{(2)}_{ij}(\mathbf{X}) &=& \sum^K_{k=1}g_{ik}(\mathbf{X})g_{jk}
                           (\mathbf{X})\label{DeqG2}
\end{eqnarray}
for $i,j=1,\dots,K$,
describing the evolution of the joint probability 
density function (PDF) $f(\mathbf{X},t)$ 
by means of the Fokker-Planck equation \cite{fpeq,gard}:
\begin{eqnarray}
\frac{\partial}{\partial t} f(\mathbf{X},t) &=&
-\sum_{k=1}^K \frac{\partial}{\partial x_k} 
     D^{(1)}_k(\mathbf{X})f(\mathbf{X},t) \cr
&  & 
+\sum_{k=1}^K\sum_{m=1}^K \frac{\partial^2}{\partial x_k\partial x_m} 
     D^{(2)}_{km}(\mathbf{X})f(\mathbf{X},t) .
\label{eq:fp}
\end{eqnarray}

As done previously in other 
contexts\cite{physrepreview,lind05,friedrich00,ghasemi07,kleinhans05,%
boettcher06,lind10}, the drift vector and the diffusion matrix can be 
derived directly from the data. 

Statistically, the drift and diffusion
coefficients coefficients $D_i^{(1)}$ and $D_{ij}^{(2)}$ are defined as
\begin{equation}
\mathbf{D}^{(k)}(\mathbf{X})=\lim_{\tau \rightarrow0}\frac{1}{\tau}
                          \frac{\mathbf{M}^{(k)}(\mathbf{X},\tau)}{k!},
\label{DefCoefKM}
\end{equation}
with the first and second conditional moments given by
\begin{subequations}
    \begin{eqnarray}
      M_i^{(1)}(\mathbf{X},\tau) &=& 
      \left\langle Y_i(t+\tau)-Y_i(t) | \mathbf{Y}(t)=\mathbf{X} \right 
        \rangle \label{M1}\\
        M_{ij}^{(2)}(\mathbf{X},\tau) &=& 
        \left\langle (Y_i(t+\tau)-Y_i(t)) \right. \nonumber \\
        && \left. \cdot (Y_j(t+\tau)-Y_j(t)) | \mathbf{Y}(t)=\mathbf{X}  \right\rangle  \label{M2} ,
        \end{eqnarray}
      \end{subequations}
Here $\langle \cdot| {\mathbf{Y}(t)=\mathbf{X}} \rangle$ symbolizes conditional averaging over all events 
      that fulfill the condition $\mathbf{Y}(t)=\mathbf{X}$. 

To determine the underlying Langevin equations,
defined in Eq.~(\ref{Lang2DVect}), one additionally  needs
to solve Eqs.~(\ref{Deqh1}) and (\ref{DeqG2}).
In particular, the calculation of matrices $\mathbf{g}$ from the 
diffusion matrices requires  to solve 
$\mathbf{D^{(2)}}=\mathbf{g}\mathbf{g}^T$, 
e.g.~by means of diagonalization,
$\mathbf{D_{diag}^{(2)}}=\mathbf{P}\mathbf{D^{(2)}}\mathbf{P}^{-1}$, 
with $\mathbf{P}$ the orthogonal matrix of eigenvectors of $\mathbf{D^{(2)}}$.
The family of solutions is given by 
$\mathbf{g}= \mathbf{P^T} \sqrt{\mathbf{D_{diag}^{(2)}}} \mathbf{P} \mathbf{O}$, 
where $\mathbf{O}$ is an arbitrary orthogonal matrix, 
obeying $\mathbf{O}\mathbf{O^T}=1$.
The matrices $\mathbf{D^{(2)}}$ are symmetric
and positive semi-definite with all their eigenvalues real and 
non-negative (see Eq.~(\ref{M2})), and therefore 
$\sqrt{\mathbf{D_{diag}^{(2)}}}$ is well-defined.
For any choice of $\mathbf{O}$ the analysis below
does not change, and therefore we choose for simplicity
$\mathbf{O}$ as the identity matrix.

The computation of the conditional moments is based on their
statistical $\tau$-dependence for small 
$\tau$\cite{physrepreview,lind10}. 
Previous works showed that Eqs.~(\ref{M1}) and (\ref{M2}) are an 
operational definition of the conditional moments that can easily
be implemented for the direct estimation of the drift and 
diffusion coefficients from the data\cite{physrepreview,lind10}. 
In some practical situations, the limit in Eq.~(\ref{DefCoefKM}) 
can be approximated by the slope of a linear fit of the corresponding 
conditional moments at small $\tau$. 
When such linear fit is not possible, an alternative estimate is to 
consider the first value of $M(\tau)/\tau$ at the lowest value
of $\tau$\cite{kleinhans05}. We will use this latter estimate for deriving the
drift and diffusion coefficients, underlying the evolution of 
\no\ concentration in Lisbon.

Within this  framework, we consider the two-dimensional system of \no\ 
concentrations ${\bf X}=(x_1,x_2)$ describing the fluctuations at 
the stations of Chelas and Avenida da Liberdade, see Fig.~\ref{fig1}.
In order to comply with a Langevin process, as defined in 
Eq.~\eqref{Lang2DVect}, we first verify that 
both data sets exhibit Markovian properties, which we show next for 
component $x_1$ only, for sake of clarity.
For $x_2$ the results are similar.

As Fig.~\ref{fig5} indicates, the conditional moments of the time series
show no evidence of measurement noise as $\tau$ approaches 
zero\cite{boettcher06}: $M^{(1)}/\tau$ do not diverge when $\tau\to 0$. 
This is true, both before and after detrending.
For $\tau$ smaller than a limiting value $\tau_c$, 
some oscillations are observed in the case without detrending, although 
they have no impact on the estimate of the corresponding
Kramers-Moyal coefficients, as compared to our method of using the value 
at $\tau=1$ as estimate. For details see Ref.~\cite{kleinhans05}.

The resulting components of the  drift and 
diffusion coefficients are plotted in Fig.~\ref{fig6}. 
As one sees all surfaces are adequately fitted by a quadratic polynomial 
\begin{equation}
  \label{eq:poly}
  p(x_1,x_2) = a_1 x_1^2 + a_2 x_2^2 + a_3 x_1 x_2 + a_4 x_1 + a_5 x_2 + a_6 \, ,
\end{equation}
where $p$ denotes the drift and diffusion components, $D^{(1)}_i$ and 
$D^{(2)}_{ij}$ respectively, and the coefficients $a_i$ are computed from 
a least-square procedure on the drift and diffusion components as 
functions of the detrended variables $x_1$ and $x_2$.

\section{Analysis of Markov properties for the series of 
         \no\ concentrations}
\label{sec:markov}

\begin{figure}[t] 
  \centering
  \includegraphics[width=0.5\textwidth]{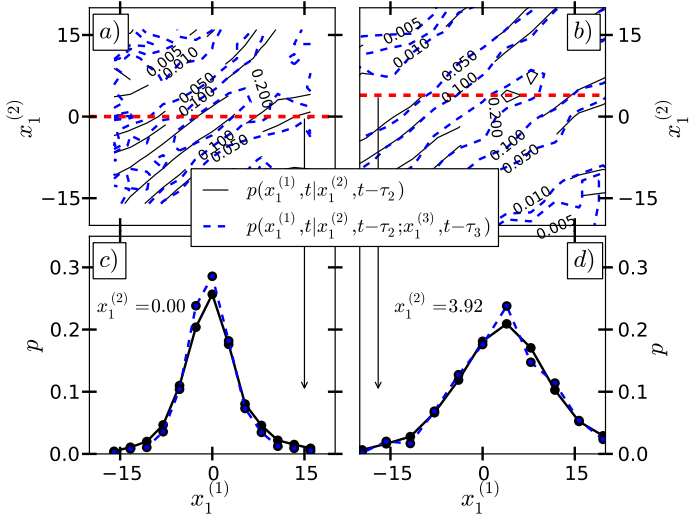}
  \caption{\protect 
      Contour plots of conditional probabilities (solid curves) and
      conditional two-point probabilities (dashed curves) computed from 
      the detrended time series $x_1$ and $x_2$ with $\tau_2=1$ hour for
      {\bf (a)} $\tau_3=2$ hours and 
      {\bf (b)} $\tau_3=10$ hours. 
      The corresponding cuts through contour planes, indicated by the
      horizontal dashed lines, are shown in
      {\bf (c)} and {\bf (d)} with a good matching between 
      the respective one-point and two-point conditional probabilities. 
      The distributions were computed with
      $13$ bins for each variable using a sample of $10^5$ data points.}
  \label{fig3}
\end{figure}
\begin{figure}[t] 
  \centering
  \includegraphics[width=0.5\textwidth]{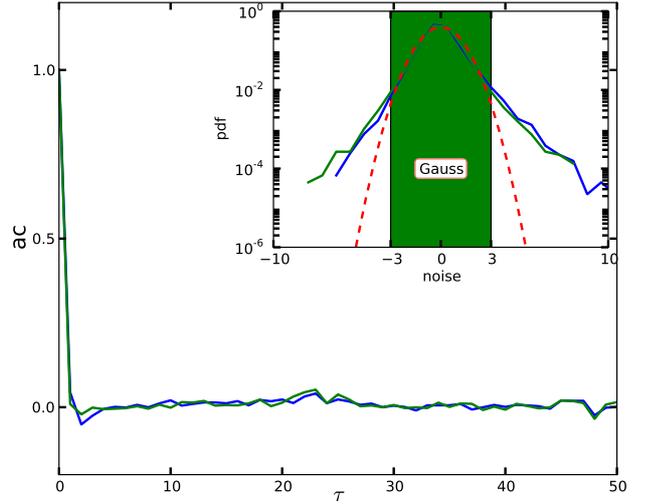}
  \caption{\protect
      Autocorrelation of the reconstructed dynamical noises 
      $\Gamma_1,\Gamma_2$ (stochastic fluctuation), indicating 
      that they are $\delta$-correlated. The inset shows the 
      probability density function (PDF) of
      the reconstructed noise normalized to variance 1 (lines) and a normal
      distribution for comparison (dashed line).}
\label{fig4}
\end{figure}

The Markovian nature of the variable $x_1$ can be investigated by 
considering the differences between the conditional one-point
probability $p(x_1^{(1)},t|x_1^{(2)},t-\tau_2)$ and the conditional 
two-point probability $p(x_1^{(1)},t|x_1^{(2)},t-\tau_2; x_1^{(3)},t-\tau_3)$. 
If the process is Markovian on time scales larger than $\tau_2$,
then these probability distributions should not differ 
significantly\cite{physrepreview} for any choice of $\tau_3$.
Indeed, as can be seen from Fig.~\ref{fig3}, the Markovian properties 
seem to be  fulfilled for $\tau_2=1$h and both $\tau_3 =  2$h and 
$\tau_3 =  10$h .
We therefore observe strong indications that the process is Markovian 
already at the sampling rate of the data points of $1$h and for time 
lags longer than $1$h.

Further, it is also necessary to check the Gaussian nature of the 
stochastic force $\Gamma$ and ascertain  it indeed  obeys 
Eqs.~(\ref{stdgamma}).
Using the measured time series and the estimated KM-coefficients,  
the noise $\mathbf{\Gamma}(t)$ can be reconstructed  from  a numerical 
discretization of Eq.~\eqref{Lang2DVect} solved with respect to
$\Gamma$\cite{reconstr}, namely solving
  \begin{equation}
    \mathbf{\Gamma} = \mathbf{g}^{-1}(\mathbf{X}) \left( \mathbf{\bar{X}} - \mathbf{h}(\mathbf{X}) \right) \, ,
    \label{gamma}
  \end{equation}
  where $\bar{\mathbf{X}}= \mathbf{X}(t+1) - \mathbf{X}(t)$ and  $\mathbf{h}(\mathbf{X})$ and
  $\mathbf{g}(\mathbf{X})$ are evaluated at $\mathbf{X}=\mathbf{X}(t)$.

The resulting 
noise is analyzed with respect to its autocorrelation, shown in 
Fig.~\ref{fig4}:
the autocorrelation decays to zero for the very first values of $\tau$,
which strongly supports to treat $\Gamma_1$ and $\Gamma_2$ as a white, 
$\delta$-correlated noise source.

For ascertaining the Gaussian nature of the stochastic sources we
plot in the inset of Fig.~\ref{fig4} the PDF of the reconstructed noise 
time series $\Gamma_1$ and $\Gamma_2$ (solid lines) against a 
Gaussian distribution (dashed lines).

As one sees from the inset, in the range comprising over 95\% of the
Gaussian noise, the distributions for the stochastic sources are well
approximated by a Gaussian distribution. 
We find it reasonable to assume, therefore,
that the data series can be approximated sufficiently well  by a 
Fokker-Planck equation.
The deviations observed for the extreme values, are common
in the analysis of long-term field measurements, showing
tails for close to exponential decay.

From the tests described in this section one may satisfactorily take 
the series $x_1$ and $x_2$
as a  set described by two coupled Langevin Equations,
Eq.~\eqref{Lang2DVect} with $K=2$. Next we derive these equations from the
sets of measurements $x_1$ and $x_2$.

\section{Deriving optimal variables:
         eigensystem for \no\ measurements at different stations}
         
\label{sec:eigensyst}

Having successfully determined the drift and diffusion constants 
describing the respective deterministic forcing and stochastic fluctuations 
of the system of \no\ concentration measurements, we now determine the 
eigensystem of the diffusion matrices and investigate its principal 
directions.  
This procedure was described in detail in \cite{vitor} and  was previously applied to a two-dimensional sub-critical 
bifurcation\cite{gradisek_eigenvectors} and to the analysis of human 
movement\cite{vanMourik_eigenvectors}.  It will 
be briefly outlined here, for $K$ variables.
 
Diffusion matrices are numerically estimated on a mesh of points in 
phase space, as shown for example in Fig.~\ref{fig6}c-e.
Then  at each mesh point the $K$ eigenvalues and corresponding 
eigenvectors  of the estimated  matrices are calculated.
The diffusion matrices contain  information about the stochastic 
fluctuations acting on the system and we use the local eigensystems of 
the matrix for a further characterization of these forces.
In particular, a vanishing eigenvalue indicates that the corresponding 
stochastic force may be neglected.

We are looking for a transform of the original coordinates 
$\mathbf{X}=\{ x_i \}$ into new ones 
$\mathbf{\tilde{X}}=\{ \tilde{x}_i \}$, such that the new coordinates are 
aligned in the directions of the eigenvectors of the diffusion matrix in 
each mesh point, i.~e.~the principal direction in which the diffusion 
matrix is diagonal. 
Diagonalizing the diffusion matrix decouples the stochastic contribution
in the set of variables, and if the eigenvalues in the transformed 
coordinates are significantly different, we are able to restrict our 
investigation to the coordinates with lower stochasticity.

We therefore look for a two-times continuously differentiable function 
$\f{F}$ with 
\begin{equation}
\f{\tilde{X}}={\f{F}}(\f{X},t),
\label{transf}
\end{equation}
for which\cite{fpeq,gard} the deterministic and stochastic
parts in the Langevin systems of equations, transform respectively 
as\cite{vitor} 
\begin{eqnarray}
\tilde{h}_i^{(1)}({\bf \tilde{X}}) &=& 
         \sum_{k=1}^N \Big ( h_k^{(1)}({\bf {X}}) \frac{\partial{F_i}} 
              {\partial{x_k}}\cr
       & & +\sum_{l=1}^N \sum_{j=1}^N g_{lj}({\bf {X}})g_{kj}({\bf {X}})
         \frac{\partial{^2F_i}}{\partial{x_k}\partial{x_l}}\Big ) 
        \label{drifty}\\
\tilde{g}_{ij}({\bf {\tilde{X}}}) &=& 
        \sum_{k=1}^N g_{kj}({\bf {X}})\frac{\partial{F_i}}{\partial{x_k}}
\label{gprime-jacobian}
\end{eqnarray}
where the second equation reads $\mathbf{\tilde{g}}({\bf {\tilde{X}}}) 
= \mathbf{J}({\bf {{X}}})\mathbf{g}({\bf {{X}}})$,
with $\mathbf{J}({\bf {{X}}})$ the Jacobian of our transformation $\f{F}$. 
For reasons of clarity in the following we do not explicitly notate the 
dependence on ${\bf {X}}$ and ${\bf {\tilde{X}}}$.

The eigenvectors $\mathbf{u}_k$ of matrices $\mathbf{\tilde{g}}$ with 
coordinates in local bases $\tilde{\mathbf{e}}_i$, can be incorporated 
in  matrices
${\f U}=[{\f u}_1 \quad {\f u}_2 \quad \dots \quad {\f u}_K] $.
Defining $\mathbf{\tilde{U}}$ as
$\mathbf{U} = \mathbf{J^T}\mathbf{\tilde{U}}$ one then obtains
(see Eq.~(\ref{gprime-jacobian}))
\begin{eqnarray}
\mathbf{\tilde{U}}^{T}\mathbf{\tilde{g}}\mathbf{\tilde{g}}^T
\mathbf{\tilde{U}} &=&
\mathbf{U}^{T}\mathbf{g}^T\mathbf{g}\mathbf{U} .
\label{diagonal}
\end{eqnarray}

By definition the inverse transform ${\f F}^{-1}(\tilde{\f X})$ is chosen 
such that the normalized eigenvectors are given by
\begin{equation}
{\f u}_k=\frac{1}{s_k}\frac{\partial{\f F}^{-1}}{\partial{\tilde x_k}}
\label{tangent}
\end{equation}
with 
\begin{equation}
s_k=\left \vert \frac{\partial{\f F}^{-1}}{\partial{\tilde x_k}} \right \vert .
\label{sk}
\end{equation}
i.e.~the respective square sum of the columns in the Jacobian of the 
inverse transform.
Taking into account this scaling factor the eigenvalues in the new 
coordinate system can be calculated\cite{vitor}, 
\begin{equation}
\tilde{\f D}^{(2)}={\rm diag}\left[\frac{\lambda_1}{s_1^2} \quad 
            \frac{\lambda_2}{s_2^2} \quad \dots \quad 
            \frac{\lambda_K}{s_K^2}\right]  ,
\label{Dtilde2}
\end{equation}
where $\lambda_i$ ($i=1,\dots,K$) are the eigenvalues of the
diagonalized matrix $\mathbf{D}^{(2)}_{diag}$.

In general, 
the eigenvalues of the diffusion matrix indicate the amplitude of 
the stochastic force and the corresponding eigenvector indicates 
the direction towards which such force acts. These directions
can be regarded as principal axes of the underlying stochastic 
dynamics\cite{vitor}.

In particular the vector field yielding at each mesh point the 
eigenvector associated to the smallest eigenvalue of matrix $\mathbf{g}$ 
defines the paths in phase space towards which the fluctuations are
minimal.
If this eigenvalue is very small compared to all the 
other, the corresponding stochastic force can be neglected and the system 
can be assumed to have only $K-1$ independent stochastic forces,
reducing the number of stochastic variables in the system.

Notice however that, whereas in a Cartesian coordinate system the 
eigenvalues are strictly related to the amplitude of diffusion in the 
corresponding eigenvectors, a nonlinear transformation  usually changes 
the metric\cite{vitor,lade09b}. 
In the transformed system the direction of the maximal eigenvalue is not 
necessarily the direction with the highest diffusion. This disparity is 
accounted for by the factor $s_i$ above.
A much more simple case occurs when the eigenvectors are parallel
(or almost) to a fixed direction, meaning that the eigendirections at 
each point in phase space are the same but rotated by a constant angle.
Next we address such situation.

\section{Transform of \no\ concentrations to the  stochastic 
         eigendirections}
\label{sec:transform}

\begin{figure}[htb]
  \centering
  \includegraphics[width=0.49\textwidth]{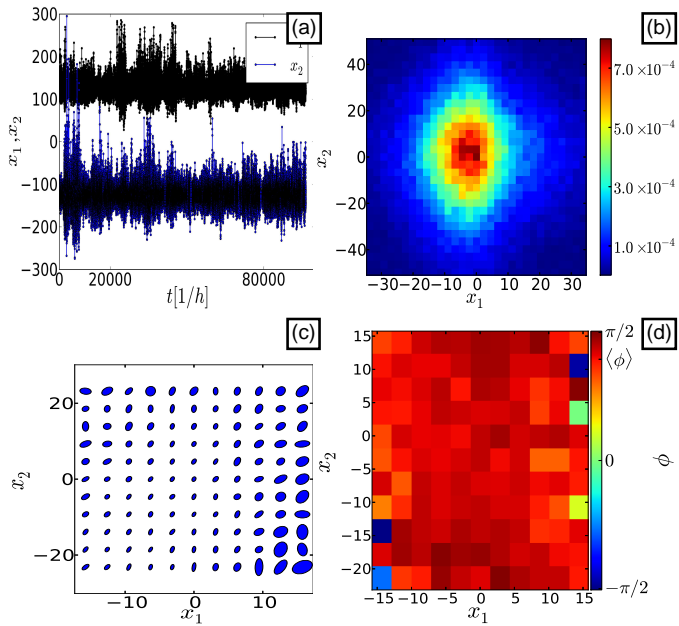}
  \caption{
    {\bf (a)} The original time-series $x_1$ and $x_2$ and 
    {\bf (b)} their joint probability density function.
    In {\bf (c)} we plot the two (orthogonal) eigenvectors
    defining the semi-axis of an ellipsoid. The major semi-axis
    is associated with the largest eigenvalue and correspondingly
    the minor semi-axis with the smallest. 
    For each grid point in phase space $(x_1,x_2)$ we plot 
    {\bf (d)} the angle $\phi$ between the eigenvector associated
    to the largest eigenvalue and the positive $x$-semi-axis.
    As one sees from (c) and (d) the angle $\phi$ does not show
    significant disparity in its values within the considered 
    range, indicating a strong uncoupling between the two 
    stations (see text). Thus, a global rotation of the axis 
    may be considered (see Fig.~\ref{fig8}).}
\label{fig7}
\end{figure}

In this section we apply the procedure described previously to the
two series of \no\ concentration, $x_1$ and $x_2$ in Chelas and
Avenida da Liberdade, shown in Fig.~\ref{fig7}a.
The joint PDF of both concentrations $x_1$ is $x_2$ is plotted in
Fig.~\ref{fig7}b showing the region in phase-space most visited
by the bivariate series $(x_1,x_2)$.
A plot of the eigenvectors of ${\f D}^{(2)}$  in Fig.~\ref{fig7}c suggests  
that a continuous and smooth description of the corresponding sorted 
eigenvalues exists. 
Here we place at each grid point of the phase space one ellipsoid whose 
major and minor axis are given by the (non-normalized) eigenvectors
associated to the largest and smallest eigenvalue respectively.
Since the two eigenvalues are different, the eigenvector corresponding to 
the lower eigenvalue describes the direction of minimum stochasticity.  
The eigenvectors and eigenvalues of the diffusion matrix give 
locally the principal directions of stochastic fluctuations (diffusion).

In general, from a plot as the one in Fig.~\ref{fig7}c it is possible
to derive numerically the variable transformation in Eq.~(\ref{transf}):
at each grid point one determines the angle between the ``largest'' 
eigenvector and the positive horizontal axis. Figure \ref{fig7}d shows the
angle $\phi$ for the bivariate series $(x_1,x_2)$. Rotating each ellipsoid
separately by the respective $\phi$-angle aligns the largest eigenvector
along the horizontal direction and the smallest eigenvector along the
vertical direction, yielding the two new (transformed) variables 
$\tilde{x}_1$ and $\tilde{x}_2$. This angle can be derived at each grid
point from the corresponding diffusion ${\f g}$ components namely
\begin{equation}
\tan 2\phi  = \frac{2g_{12}}{g_{11} -g_{22}}
\label{phi}
\end{equation}
The angle $\phi$ or its absolute value quantifies the relative off-diagonal 
contribution that describes the coupling of the noise terms by the diffusion 
matrix.
\begin{table}[tb]
  \centering
  \footnotesize
    \begin{tabular}{|l|c|c|c|c|} \hline
     &$y$ &$x$ &$\mathbf{\tilde{x}}$ &$\hat{x}$ \\ \hline
    $\langle \rangle_1$ &$36$ &$5.82\cdot 10^{-6}$ &$0.00935$ &$-0.0145$\\ \hline
    $\langle \rangle_2$ &$64.7$ &$5.22\cdot 10^{-6}$ &$0.0541$ & $-0.0721$\\ \hline
    $\sigma_1$  &$26.8$  &$17.3$  &$26.7$  &$18.3$\\ \hline
    $\sigma_2$  &$40.8$  &$25.5$  &$15.1$  &$23.3$\\ \hline
    $\langle \left| \phi \right| \rangle $               &$1.11$$\pm$$0.282$      &$1.31$$\pm$$0.203$      &$0.117$$\pm$$0.114$ &$1.33$$\pm$$0.105$    \\ \hline
    $\langle Q \rangle$               &$0.703$$\pm$$0.182$       &$0.714$$\pm$$0.103$      &$0.679$$\pm$$0.0946$  &$0.593$$\pm$$0.0445$\\ \hline
    $\langle R_1 \rangle $            &$0.157$$\pm$$0.125$    &$0.247$$\pm$$0.167$    &$0.19$$\pm$$0.122$ &$0.169$$\pm$$0.118$ \\ \hline
    $\langle R_2 \rangle $             &$0.183$$\pm$$0.13$    &$0.207$$\pm$$0.134$    &$0.245$$\pm$$0.15$  &$0.243$$\pm$$0.153$ \\ \hline
    $\mu$              &$0.479$      &$0.457$     &$-0.254$    &$-0.401$\\ \hline
    \end{tabular}
  \normalsize
  \caption{
    Characterizing different pairs of variables:
    the original pair of measures $\mathbf{y}$,
    the detrended pair of variables $\mathbf{x}$,
    the transformed pair $\mathbf{\tilde{x}}$, and, for comparison, the pair $\mathbf{\hat{x}}$ transformed according to the simpler rules in \eqref{eq:triv_transf}.
    For each variable or pair of variables ($i=1,2$) we show the
    mean $\langle\rangle_i$ and standard deviation $\sigma_i$ of 
    their distribution of observed values together with the rotation
    angle $\phi$ averaged over phase space, as well as the average 
    coefficients $Q$ and $R_i$ for evaluating their stochastic and 
    deterministic contributions.
    See Eqs.~(\ref{phi}), (\ref{Q}) and (\ref{R}).
    The correlation coefficient between both variables is also
    given in each case (see text).}
  \label{tab4}
\end{table}

In general, what does 
such a transformation add to our understanding about the system?
First, by definition the transformation decouples independent
stochastic forces in the system.
The original (detrended) pair of variables as well as the transformed
pair of variable obey Eq.~(\ref{Lang2DVect}), with one important difference:
the transformed pair of variables are such that each variable has a
stochastic contribution governed by one independent stochastic force 
alone. In other words ${\f g}(x_1,x_2)$ is diagonal.
For the original pair of variables the stochastic contribution
mixes both independent stochastic forces.
Second, in a reference frame 
where the two independent stochastic forces are decoupled, their 
minimum and maximum magnitude reach the largest difference between them.
In other words, one aligns the major and minor axis of the ``diffusion
ellipsoids'' shown in Fig.~\ref{fig7}c.
In the particular case when one of the magnitudes is much smaller than 
other one, one of the variables can even be disregarded
as a stochastic variable, reducing the number of stochastic variables
describing the system. A generalization to $K$ variables is 
straightforward.

\begin{figure*}[htb]
  \centering
  \includegraphics[width=1.0\textwidth]{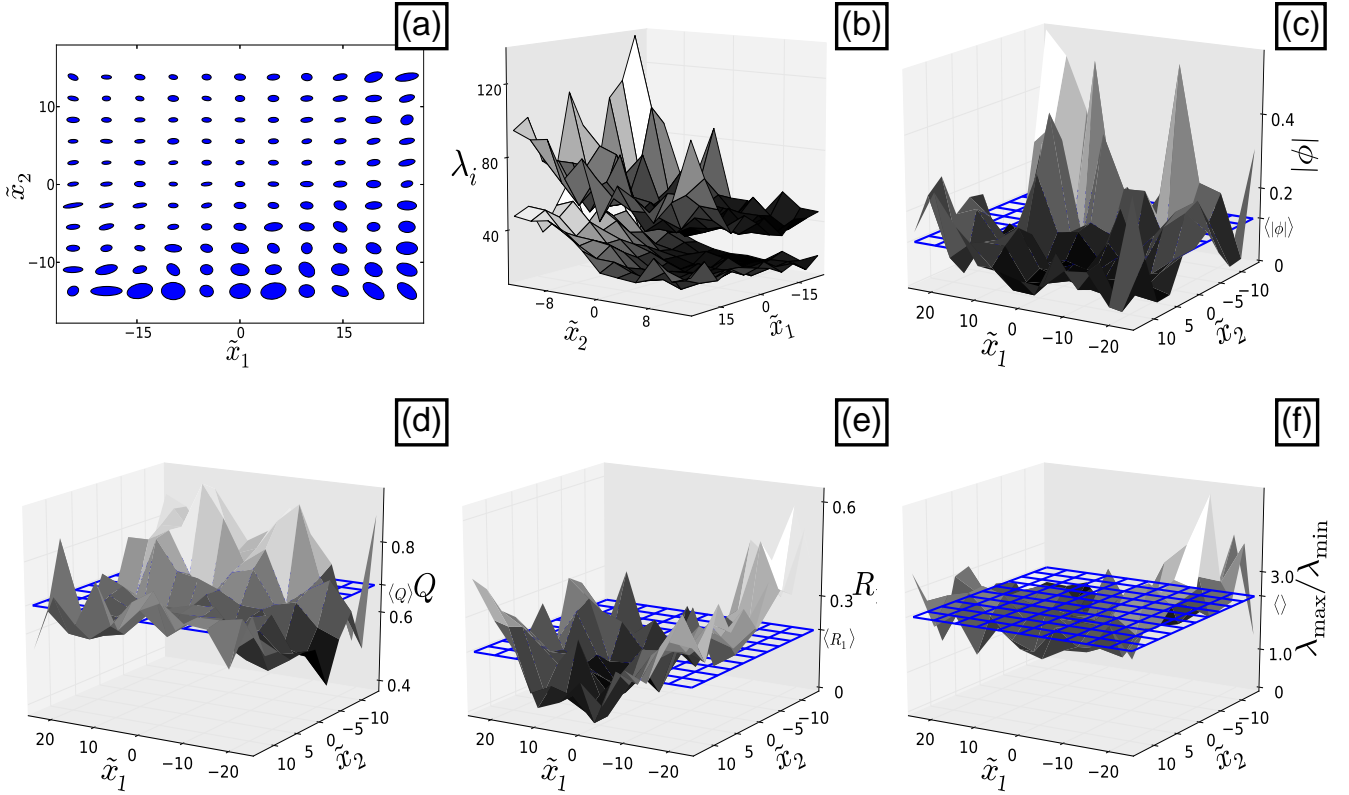}
  \caption{\protect 
    Transformed variables through a global rotation of the $x_1$
    and $x_2$ axis (check Fig.~\ref{fig7}) into new variables,
    $\tilde{x}_1$ and $\tilde{x}_2$.
    {\bf (a)} Eigenvectors of the transformed variables and
    {\bf (b)} its eigenvalues derived for the diffusion matrix
              of the transformed variables, together with 
              quantities to evaluate some underlying properties
              of the system, namely
    {\bf (c)} the rotation angle $\phi$ (see Eq.~(\ref{phi})),
    {\bf (d)} the asymmetry of the stochastic influence at each 
              variable, given by $Q$ in Eq.~(\ref{Q}),
    {\bf (e)} the deterministic coefficient $R$ in Eq.~(\ref{R}) and
    {\bf (f)} the quotient between the largest $\lambda_{max}$ and the
              smallest $\lambda_{min}$ for each grid point in the
              transformed phase position.
              For each property the average value is showed with
              horizontal surfaces and an explicit indication at the 
              vertical axis.}
\label{fig8}
\end{figure*}

The value of $\langle\phi\rangle$ shown in Tab.~\ref{tab4} is
the $(x_1,x_2)$-averaged angle $\langle\phi\rangle\simeq 0.40\pi\sim
\pi/2$ indicated at the scale of \ref{fig7}d.
Further, one inspection of Fig.~\ref{fig7}c and \ref{fig7}d
enables the observation that in our present case 
the $\phi$-angle remains approximately constant at any grid point. 
Similar
observations were made for the other pairs of stations in Lisbon (not shown).
Consequently, we may conclude that for our set of stations a global
rotation is enough to align the ``diffusion ellipsoids''.
For the stations in Chelas and Avenida da Liberdade,
Figure \ref{fig8}a shows the result obtained after 
performing a global rotation by the median
$\mathrm{median}( \phi) \simeq 0.43\pi$.

In Fig.~\ref{fig8}b both eigenvalues $\lambda_i$ are plotted,
corresponding to the length of the major and minor axis of the diffusion
ellipsoids. While there is a significant difference between both
eigenvalues, $\lambda_{max}\sim 2.5\lambda_{min}$ as shown in Fig.~\ref{fig8}f, 
they are of the same order of magnitude. Such observation indicates the
presence of two independent stochastic forces driving the bivariate
signal $(x_1,x_2)$.

The stochastic contribution for each variables of the pair $(x_1,x_2)$ 
obeying Eq.~(\ref{Lang2DVect}) can be compared through one
parameter $Q$ defined at each point $\mathbf{x}$ in phase-space as
\begin{equation}
Q^2(\mathbf{x}) = \frac{g_{11}^2(\mathbf{x})+g_{12}^2(\mathbf{x})}
                       {g_{21}^2(\mathbf{x})+g_{22}^2(\mathbf{x})}
\label{Q}
\end{equation}
where one orders the rows of matrix $\mathbf{g}$ to guarantee
$Q<1$, i.e.~variable $x_1$ is chosen as the one having lower
stochastic contribution. When $Q=1$ both stochastic contributions
are equal. When $Q\ll 1$ one stochastic contribution 
can be neglected, reducing by one the number of stochastic 
contributions in the system. 
For an arbitrary number of stochastic variables, the generalization
of Eq.~(\ref{Q}) is straightforward\cite{vitor}.

Table \ref{tab4} shows the value of coefficient $Q$ for
the set of measurements $\mathbf{y}$, for the detrended variables
$\mathbf{x}$ and for the transformed detrended variables 
$\mathbf{\tilde{x}}$.
The coefficient is averaged over the sample of points in the
corresponding phase space.
For $\mathbf{y}$ and $\mathbf{x}$ the smallest stochastic
contribution has a magnitude of approximately $70\%$ of the largest
one, while for the transformed variables it decreases more than $2\%$. 
This magnitude is not small enough to permit neglecting one variable.
We consider this finding the central result of this letter:
before transformation the pair of detrended variables include already
two independent stochastic forces of the same order of magnitude.

One note is however important to stress at this point.
The method applied here to empirical data deals with a transformation
that operates on the diffusion matrix alone. No constraints related
to the drift functions, $h_1$ and $h_2$ are considered. 
To evaluate the predictability of each variable $i$ one needs to
compare the total amplitude of the stochastic term with the
deterministic term, namely
\begin{equation}
R_i^2(\mathbf{x})=
\frac{h_i^2(\mathbf{x})}{g_{i1}^2(\mathbf{x})+g_{i2}^2(\mathbf{x})} .
\label{R}
\end{equation}
Such expression is also straightforwardly extended to $K$ variables.
The larger $R_i$ the more predictable the variable $i$ may be, i.e.~the
smaller the stochastic overall contribution is compared to the
deterministic part governing the evolution of the variable.
In our present case, as given in Tab.~\ref{tab4}, while the
detrending $y\to x$ of our measurements increases the predictability
of the non-periodic modes in time, the global rotation has no 
major effect: both coefficients
$R_i$ maintain the same order of magnitude after transform.

The correlation coefficient $\mu$ between both stations is also given in
Tab.~\ref{tab4}. While detrending has no significant effect on the
correlation, the transform $x_i\to\tilde{x}_i$ indeed decreases its
absolute value.

Figure \ref{fig8}c, \ref{fig8}d and \ref{fig8}e illustrate the numerical
result of each property, $\phi$, $Q$ and $R$ for the transformed variables.
Similar to such variables is the quotient between the maximum and minimum
eigenvalues, shown in Fig.~\ref{fig8}f. Similar plots are obtained for
the other possible pairs of stations.

\section{Comparison with standard methodologies}
\label{sec:compari}

In this Section we first address the question of how good the coordinate 
transform derived above is compared to other, possibly simpler 
transforms.

For example, we may consider a transform to coordinates which 
describe the mean value and difference between the two measured time 
series, e.g.
\begin{eqnarray}
  \label{eq:triv_transf}
   \begin{pmatrix} \ttxa \\ \ttxb   \end{pmatrix}  &=& \tfrac{1}{2}  
\begin{pmatrix} 
x_1 + x_2 \\ x_1 - x_2 \end{pmatrix} \,.  
 \end{eqnarray}
This choice is the simplest one for two variables, one describing
the total amount $x_1+x_2$ and another describing the relative amount
$x_1-x_2$.
For such choice of variables we obtain a value of $Q=0.59$, which is 
essentially the same as for our ``optimized'' variables (see Tab.~\ref{tab4}), 
The absolute value of the angle $\langle \left| \phi \right| \rangle$ is however considerably larger than for our optimized transform, as is the correlation coefficient between the time series, meaning that this simple transform fails to decouple the noise sources. 
The drift-diffusion quotients yield
$R_1=0.17$ and $R_2=0.24$, showing again no better predictability
in comparison with the original variables.

In our case we saw that the eigendirections do not depend much on the
detrended variables $x_1$ and $x_2$, which implies that they are functionally 
decoupled. 
However, sometimes it is necessary to consider a proper scaling of the 
variables\cite{vitor}.  
In such cases, we find it advisable to use a more general
transform to generalized polar coordinates given by
\begin{eqnarray}
  \label{eq:gen_rot}
 \begin{pmatrix} \txa \\ \txb   \end{pmatrix}  &=&  \begin{pmatrix} r_g 
\\ \theta_g \end{pmatrix}  \nonumber \\
 &=&  \begin{pmatrix} \sqrt{ (\alpha x_1+\beta)^2 + (\gamma x_2 +\delta)^2) } \\ \arctan \left( \frac{\gamma x_2 + \delta}{\alpha x_1 +\beta} \right) +\epsilon   \end{pmatrix} 
\end{eqnarray}
where in general 
the radial and angle variables, $r_g$ and $\theta_g$, 
are functions of the detrended variables $x_1$ and $x_2$.
This approach has the advantage that the inverse transform ${\f F}^{-1}$ 
is given by the simple form
\begin{equation}
  \label{eq:eq:gen_rot_inverse}
    \begin{pmatrix} 
x_1 \\ 
\ \\
x_2   
    \end{pmatrix} =   
    \begin{pmatrix} 
    \frac{1}{\alpha} [ r_g \cos (\theta_g-\epsilon) -\beta] \\ 
    \ \\
    \frac{1}{\gamma} [r_g \sin (\theta_g-\epsilon) -\delta]  \end{pmatrix}  \,.
\end{equation}
In addition, the metric factors introduced by the polar transform can result 
in a more pronounced separation of the eigenvalues in the transformed 
coordinates.

The other question
addressed within this section is the comparison between 
methods applied to choose the most appropriate inputs. Theoretically, any 
set of input data can be fed into a model for training 
and evaluation. However, the number of possible variables to be used and 
the number of ways they can be presented is too diverse to test all possible 
combinations. A number of statistical methods can be applied in 
order to choose the most appropriate set of predictors or inputs.
Examples are, among other, stepwise regression, PCA, cluster analysis and
ARIMA. 
For details, see Ref.~\cite{Wilks2006} and references therein. 
Such pre-processing procedures, reduce the number of input variables into 
the models, thus eliminating the redundant information.
In these standard procedures, the selection of variables is usually made 
independently for each monitoring station. 

Another possible way to tackle redundancy is the pre-processing of data 
consisting of the computation of backward stepwise regressions (BSR) 
conducted between a target variable and all the other data sets. Based on 
the available common period data sets, one constructs a collection of 
records, composing the input vector, which includes the meteorological 
variables, air pollutant concentrations, etc, and together with it assumes
the corresponding target, which in our case is the atmospheric concentration 
of a certain pollutant.  
Subsequently, one retains the smallest subset of statistically significant 
variables to predict a certain pollutant concentration automatically at a 
given monitoring station. 
In addition, BSR allows the determination of the best time lags for each 
input variable, typically daily and weekly cycles.

The referred techniques also allow the comparison between the original 
data sets and surrogate data sets including only the  stochastic component. 
The stochastic component may be determined through a rough approximation of 
a mathematical function (e.g., sin x), or, for example, by the presented 
framework.  After the selection of variables and the determination of cyclic 
and stochastic behaviors on each time series, linear and non-linear models can 
be applied in order to model air pollution in each monitoring station. 
The forecasting capabilities of the different approaches can then be compared.
Such models are also applied to each decoupled time-series in order to 
predict next days air quality at each monitoring station.
The applications of this framework, however, allows to determine the 
stochastic component on a efficient manner, enhancing air quality predictions.

\section{Discussion and Conclusion}
\label{sec:results}

In this paper, we investigated the stochastic properties of a
set of two simultaneous series, obtained by introducing a proper 
detrending of \no\ measurements, which is able to remove periodic 
modes in the series. We focused in the measurements at two different 
stations out from a set of 22 stations in Lisbon.

Based on validity tests we assumed, that the time series after
detrending were properly modeled by a system of Langevin equations. 
The validity of this assumption is discussed in section \ref{sec:method}, 
showing that the data sets obey the Markov property to a sufficient extent.
The stochastic fluctuations show good resemblance with 
$\delta$-correlated Gaussian noise.

Calculating the 
eigenvalues of the diffusion matrices, we found a transform that leads to 
a description in which the diffusion matrices are diagonal.
Since the transformed variables are derived directly from the
transformation that diagonalizes the diffusion matrices, they correspond
to the orthogonal directions in phase space in which fluctuations
are stronger (larger eigenvalue) and weaker (smaller eigenvalue)
respectively.

Comparison between original and transformed variables showed that
the two detrended variables are driven by stochastic forces almost
decoupled from each other, showing an almost constant rotation angle
of the ``diffusion ellipsoid'' at each point of phase space.
Further, both stochastic sources have amplitudes of the order of the
deterministic terms, indicating a short horizon of predictability.
This procedure worked out well for the \no\ data, since the 
transformation of variables resulted in decoupling 
the diffusion components in the new coordinates. Other transformation
could be considered.
For instance, we discussed how this approach could be applied for other
data sets in which the diffusion ellipsoids do not align in phase
space, but instead depend in non trivial functional of the variables.
In this case, the transform maps the detrended variables into two 
polar-like coordinates.

One question that should be addressed in a forthcoming study
is to present a systematic overview on all pairs of stations studied
by us in this scope but not shown thoroughly, since it was out
of our main purposes. Doing that one would be able to compare in detail
the results obtained through the method applied in this paper with
standard methods used for forecasting \no\ concentration 
at a specific spot in the city of Lisbon.

\section*{Acknowledgments}

The authors thank DAAD and FCT for financial support through 
the bilateral cooperation DREBM/DAAD/03/2009.
FR (SFRH/BPD/65427/2009) and PGL ({\it Ci\^encia 2007})
thank Funda\c{c}\~ao para a Ci\^encia e a Tecnologia
for financial support, also with the support 
Ref.~PEst-OE/FIS/UI0618/2011.


\bibliographystyle{elsarticle-num}


\end{document}